# A NON-COOPERATIVE GAME THEORETICAL APPROACH FOR POWER CONTROL IN VIRTUAL MIMO WIRELESS SENSOR NETWORK

R.Valli and P.Dananjayan

Department of Electronics and Communication Engineering,
Pondicherry Engineering College, Pondicherry, India.
**pdananjayan@rediffmail.com**, **pdananjayan@pec.edu**

## ABSTRACT

*Power management is one of the vital issue in wireless sensor networks, where the lifetime of the network relies on battery powered nodes. Transmitting at high power reduces the lifetime of both the nodes and the network. One efficient way of power management is to control the power at which the nodes transmit. In this paper, a virtual multiple input multiple output wireless sensor network (VMIMO-WSN) communication architecture is considered and the power control of sensor nodes based on the approach of game theory is formulated. The use of game theory has proliferated, with a broad range of applications in wireless sensor networking. Approaches from game theory can be used to optimize node level as well as network wide performance. The game here is categorized as an incomplete information game, in which the nodes do not have complete information about the strategies taken by other nodes. For virtual multiple input multiple output wireless sensor network architecture considered, the Nash equilibrium is used to decide the optimal power level at which a node needs to transmit, to maximize its utility. Outcome shows that the game theoretic approach considered for VMIMO-WSN architecture achieves the best utility, by consuming less power.*

## KEYWORDS

*Wireless Sensor Network,   Power Control,   Virtual MIMO,   Non-Cooperative Game*

## 1. INTRODUCTION

The birth of wireless sensor network (WSN) has brought out the practical aspects of pervasive computing and networking. A wireless sensor network is a group of specialized sensor nodes each of which is small, lightweight and portable with a communication infrastructure intended to monitor and record conditions at diverse locations. Generally sensors are battery powered and have feeble data processing capability and short radio range [1]. The resource-constraint temperament of WSNs in terms of their size, cost, weight and lifetime [2] is a key area of apprehension for most applications using WSN. The best part of this resource constraint nature makes wireless sensor networks to be used in the context of high end applications, security applications and consumer applications [3]. The worst part of this calls for research on power limited capability thereby prolonging the network reliability and operation.

Energy efficiency and achieving reliability is a key issue in wireless sensor networks. Battery capacity is limited and it is usually impossible to replace them. Any operation performed on a sensor consumes energy, involving discharge of battery power. Hence battery power efficiency is a critical factor while considering the energy efficiency of WSN. The three domains of energy consumption in a sensor are sensing, data processing and data communication, out of which communication is the main consumer of energy. Hence transmission at optimal power level is very essential. Optimal transmit power level implies the power level which reduces the interference, increases the successful packet transmission and provides the desired quality of service. Maintaining the transmit power under control is furthermore favorable to decrease the packet collision probability, which if not leads to more retransmitted packets wasting even more energy. Hitherto energy efficiency has been investigated extensively and various approaches to achieve an energy efficient network includes, scheduling sensor nodes to alternate between energy-conserving modes of operation, competent routing algorithms, clustering, incorporating astuteness and use of spatial localization at every node to lessen transmission of redundant data. Nodes specialization to different roles such as idle, sensing, routing and routing/sensing to maximize the utility of the nodes [4] has been proposed. An approach for optimizing transmit power for an ad-hoc network scenario [5] where all the nodes uses a uniform transmit power, and numerical results of transmit power sufficient to satisfy the network connectivity has been proposed. In recent years there has been a growing interest in applying game theory to study wireless systems. [6], [7] used game theory to investigate power control and rate control for wireless data. In [8], the authors provide motivations for using game theory to study communication systems, and in particular power control. Distributed iterative power control algorithms have been proposed for cellular networks; these algorithms examine to find the power vector for all the nodes that minimizes the total power with good convergence [9], [10]. Decentralized, game theoretic adaptive mechanisms, which can be deployed to manage sensor activities with low coordination overhead has been explained [11].

In recent years, power control has received deep interest for cellular radio systems and ad-hoc wireless networks. In [12], a power control method is described as a Markov chain. An interesting characterization of power control algorithms from a control theoretic perspective can be found in [13]. A move towards node energy conservation in sensor network is cooperative multi input-multi output transmission technique [14] has been proposed and analyzed. Virtual multiple input multiple output (VMIMO) based WSN model requires sensor cooperation. If some individual sensors cooperate with others for transmission and reception, a virtual MIMO can be constructed in an energy efficient way for wireless sensor networks. The best modulation and transmission strategy to minimize the total energy consumption required to send a given number of bits is analyzed [15]. In this paper, a game theoretic approach to regulate the transmit power level of the nodes in a VMIMO-WSN is considered and investigated. The concept of game theory has been used in networks for designing mechanisms to induce desirable equilibria both by offering incentives and by punishing nodes [16-19].

The rest of the paper is organized as follows. Section 2 deals with the system model of VMIMO-WSN. In Section 3, the basics of game theory and its application in sensor networks is discussed. A non-cooperative power control game is constructed and a utility function suitable for VMIMO-WSN is designed. Simulation results are given and discussed in section 4. Finally, conclusion of the work is given in Section 5.

## 2. VIRTUAL MIMO BASED WIRELESS SENSOR NETWORK

The information theoretic predictions on large spectral efficiency of multiple-input-multiple-output (MIMO), has motivated a great amount of research in various MIMO techniques for wireless communication. As the use of MIMO technology in wireless communication grows, MIMO interference systems have engrossed a great deal of attention. [20], [21] studied the interactions and capacity dependencies of MIMO interference systems and [22], [23] explored methods for power management and interference avoidance in MIMO systems. However, a drawback of MIMO techniques is that they require intricate transceiver circuitry and huge sum of signal processing power ensuing in large power consumptions at the circuit level. This fact has prohibited the application of MIMO techniques to wireless sensor networks consisting of battery operated sensor nodes. And also nodes in a wireless sensor network may not be able to accommodate multiple antennas. Due to the circuit complexity and obscurity of integrating separate antenna, virtual MIMO concepts are applied in wireless sensor networks for energy efficient communication to hoard energy and enhance reliability. Jayaweera [24] analyzed the consumed energy of a sensor node which employs a MIMO transceiver. Dai and Xiao [25, 26] proposed cooperative MIMO systems and the use of V-BLAST techniques as a more power efficient scheme. The V-BLAST scheme there is no joint encoding requirement at the sensor nodes. The optimum time management and power budget allocation for virtual MIMO is proposed [27] and the analysis of this shows that virtual MIMO functions like actual MIMO for low signal to noise ratio.

The concept of VMIMO is explained in Figure 1 which shows the scenario of using three transmitters and two receivers. The sender node, S, transmits a message to the destination node, D. First, S transmits the message to three transmitter nodes, t1, t2, and t3. These transmitter nodes transmit the message to the receiver nodes, r1 and r2. Then, the receiver nodes forward the message to the destination node, D.

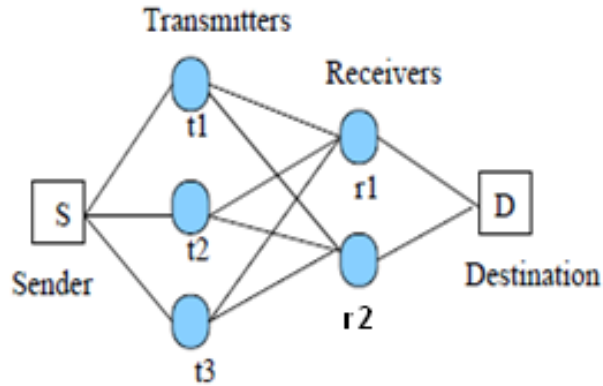

Figure 1. Virtual MIMO scenario

The total power consumption along a signal path in a VMIMO can be alienated into two main components: the power consumption of all the power amplifiers $P_{PA}$ and the power consumption of all other circuit blocks $P_C$.

The total power consumption of the power amplifiers can be approximated as

$$P_{PA} = (1+\alpha)P_{out} \qquad (1)$$

where $P_{out}$ is the transmit power which depends on the Friss free space transmission, $\alpha = \xi/\eta - 1$ with $\eta$ being the drain efficiency of the RF power amplifier and $\xi$ being the peak-to-average ratio (PAR) that depends on the modulation scheme and the constellation size.

The total circuit power consumption for a VMIMO is estimated as

$$P_C \approx N_T\left(P_{DAC}+P_{mix}+P_{filt}\right)+2P_{synth}+N_R\left(P_{LNA}+P_{mix}+P_{IFA}+P_{filr}+P_{ADC}\right) \qquad (2)$$

where $P_{DAC}$, $P_{mix}$, $P_{filt}$, $P_{synth}$, $P_{LNA}$, $P_{IFA}$, $P_{filr}$ and $P_{ADC}$ are the power consumption values for the D/A converter (DAC), the mixer, the active filters at the transmitter side, the frequency synthesizer, the low noise amplifier (LNA), the intermediate frequency amplifier (IFA), the active filters at the receiver side and the A/D converter (ADC).

The total power consumption $P_{total}$ and is given by,
$$P_{total} = P_{PA} + P_C \qquad (3)$$

If the transmission power of each sending node in a single input single output network is $P_{total}$, the transmission power of each sending node in $N_T \times N_R$ network will be $P_{total}/\min(N_T \times N_R)$,

## 3. GAME THEORY FOR SENSOR NETWORKS

Game theory is a theory of decision making under conditions of uncertainty and interdependence. In game theory, behaviour in strategic situation in a mathematic mode is captured. A strategic game consists of: a set of players, which may be a group of nodes or an individual node, a set of actions for each player to make a decision and preferences over the set of action profiles for each player. In any game utility represents the motivation of players. Applications of game theory always attempt to find equilibriums. If there is a set of strategies with the property that no player can profit by changing his or her strategy while the other players keep their strategies unchanged, then that set of strategies and the corresponding payoffs constitute the Nash equilibrium.

Game theory offers models for distributed allocation of resources and thus provides a way of exploring characteristics of wireless sensor networks. Energy harvesting technologies essential for independent sensor networks using a non cooperative game theoretic technique [28] is analyzed. Nash equilibrium was projected as the solution of this game to attain the optimal probabilities of sleep and wake up states that were used for energy conservation. The energy efficiency problem in wireless sensor networks as the maximum network lifetime routing problem is looked upon [29]. Here the transmit power levels is adjusted to just reach the anticipated next hop receiver such that the energy consumption rate per unit information transmission can be reduced.

### 3.1. Non-Cooperative Game for Power Control

Even though achieving agreeable QoS is crucial for users, they may not be willing to achieve it at arbitrarily high power levels, because power is itself a valuable resource. This motivates a reformulation of the entire difficulty using concepts from microeconomics and game theory. In this section, we will use such a reformulation to develop a mechanism for power control. The goal of this work is to control the total transmission power consumption of the sensor nodes in the VMIMO-WSN. The game is considered when source node 'i' is transmitting to destination node 'j'. A strategic game is considered which is a model of interacting decision-makers. In recognition of the interaction, the decision-makers are referred as players. Each player has a set of possible actions. The model captures interaction between the players by allowing each player to be affected by the actions of all players, and not only by his or her own action.

The existence of some strategy sets $p_1, p_2, \ldots p_{N+1}$ for the nodes $1, 2, \ldots N+1$ is assumed. These sets consist of all possible power levels ranging from the minimum transmit power $p_{min}$ to maximum transmit power $p_{max}$. In this game, if node 1 chooses its power level $p_1$, and node 2 chooses its power level $p_2$, and so on, then,

$$p = \{p_1, p_2 \cdots p_{N+1}\} \qquad (4)$$

This vector of individual strategies is called a strategy profile. The set of all such strategy profiles is called the space of strategy profiles P′. The game is played by having all the nodes concurrently pick their individual strategies. This set of choices results in some strategy profile p$\varepsilon$ P′, and is called as the outcome of the game. At the end of an action, each node i$\varepsilon$I receives a utility value,

$$u_i(p) = u_i(p_i, p_{-i}) \qquad (5)$$

$p_{-i}$ is the strategy profile of all the nodes but for the $i^{th}$ node.
The utility to any one node depends on the entire strategy profile. During every game, the node decides whether to transmit or not, rise or lower its power level, and chooses a power level if it decides to transmit. The $i^{th}$ node has control over its own power level $p_i$ only, and the utility if a node is transmitting is given as [30]

$$u_i(p_i, p_{-i}) = \frac{br}{Fp_i} f(\gamma_i) \qquad (6)$$

where,
b   is the number of information bits in a packet of size F bits
r   is the transmission rate in bits/sec using strategy $p_i$
$f(\gamma_j)$ is the efficiency function which increases with expected signal to noise ratio (SNR) of the receiving node.

The efficiency function, is defined as $f(\gamma_j) = (1-2P_e)^F$

Where, $P_e$ is the bit error rate (BER) and it is a function of SNR. With a noncoherent frequency shift keying (FSK) modulation scheme, $P_e = 0.5e^{\frac{-\gamma_j}{2}}$, with a differential phase shift keying (DPSK) modulation scheme $P_e = 0.5e^{-\gamma_j}$, and with a binary phase shift keying (BPSK) modulation scheme, $P_e = 0.5e^{\sqrt{\gamma_j}}$.

where $\gamma_j$ denotes the expected SNR of node j.
It is assumed that the utility value obtained by a node when it decides not to transmit is 0.

For a VMIMO-WSN, the net utility is given by

$$u_i(p_i, p_{-i}) = \sum_{j=1}^{N_R} \frac{br_j}{F \sum_{i=1}^{\min(N_T, N_R)} \frac{p_i}{\min(N_T, N_R)}} f(\gamma_j) \qquad (7)$$

where, $N_T, N_R$ are the number of cooperative sensors which act as VMIMO antennas.

The net utility is obtained by considering the penalty incurred by a node. The penalty incurred accounts for the energy drained by the nodes with the usage of transmission power. If the strategy of the ith node is to transmit at signal power p∈ P′, the cost incurred is a function of $p_i$, which is denoted as $A(p_i)$. $p_i$ is a random variable denoting transmitting signal power of $i^{th}$ node.

$$A(p_i) = k \times p_i \qquad (8)$$

where k is the scaling factor.

The net utility
$$u_i^{net} = \begin{cases} u_i(p_i, p_{-i}) - A(p_i) & \text{if transmitting} \\ 0 & \text{if not transmitting} \end{cases} \qquad (9)$$

A node cannot transmit at arbitrarily high power and must make a decision on a maximum threshold power $p_t$. Exceeding this threshold will bring in non beneficial net utility for the node. A node transmits at a power level $p_i$ such that $0 < p_i \le p_t$. As far as Nash equilibrium point is concerned, the expected net utility for transmitting and for being silent should be equal at the threshold, i.e., $p_i = p_t$.

$$p_s = 1 - (1-p_e)^F \qquad (10)$$

where, $p_s$ is probability of successful transmission of a packet containing F bits from node 'i' to node 'j'.

## 4. RESULTS

For performance evaluation, it is assumed that a source node is transmitting data to a destination node. The destination node not only hears from source but also from other neighbouring nodes if they are transmitting. If $\gamma_j$ is the SINR alleged by destination, then the bit error probability for the link is given by some inverse function of $\gamma_j$. The bit corruption is assumed to be independently and identically distributed. The simulation was carried in MATLAB 7.8 and the simulation parameters are given in Table.1. The performance of the proposed VMIMO-WSN using game theoretic approach is evaluated in terms of net utility and power efficiency, for various power levels and varying channel conditions.

Table 1. Simulation Parameters

| Simulation Parameter | Description |
|---|---|
| Transmission Power $(p_{min}, p_{max})$ | 1mW, 100mW |
| Signal to interference noise ratio (SINR) | -15dB to 15dB |
| Number of information bits per frame (b) | 32 bits |
| Number of bits per frame (F) | 40 bits |
| Modulation | BPSK, DPSK, FSK |
| Data Transmission rate | 1 Mbps |
| Number of transmitting and receiving antennas ($N_T$, $N_R$) | 2,2 |

## 4.1 Average Probability of Error

In the receiver side the bit error rate is affected by transmission channel noise, interference, distortion and fading. Figure 2 gives the average probability of error for different values of SINR(dB) alleged by node j.

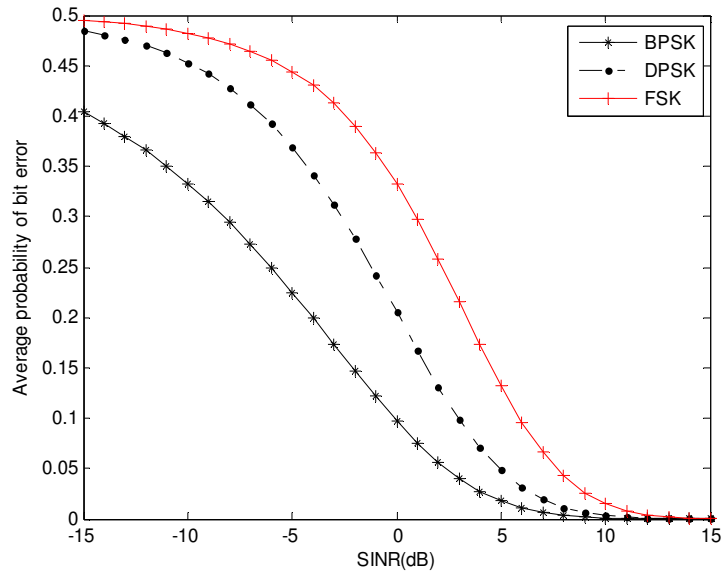

Figure 2. Average probability of bit error

The result shows that with improvement in channel condition, the average bit error rate decreases i.e., in all the cases, the average probability of error decreases monotonically with SINR. DPSK and FSK exhibit higher error rates compared to BPSK which makes BPSK appropriate for low-cost passive transmitters. This feature enables BPSK to be used as the modulation scheme in IEEE 802.15.4, 868–915 MHz frequency band.

### 4.2 Probability of Successful Frame Transmission

The primary concern of power control problem is to achieve a maximum of frame success rate, while inducing minimal power consumption. The probability of successful frame transmission is derived by considering the channel conditions and the modulation scheme. Figure 3 shows the frame success probability for different values of SINR(dB) perceived by node j. The results show that with increase in SINR, the average bit error rate decreases which in turn increases the probability of successful transmission. DPSK and FSK incur a higher probability of bit error compared to BPSK, which in turn leads to higher probability of frame error.

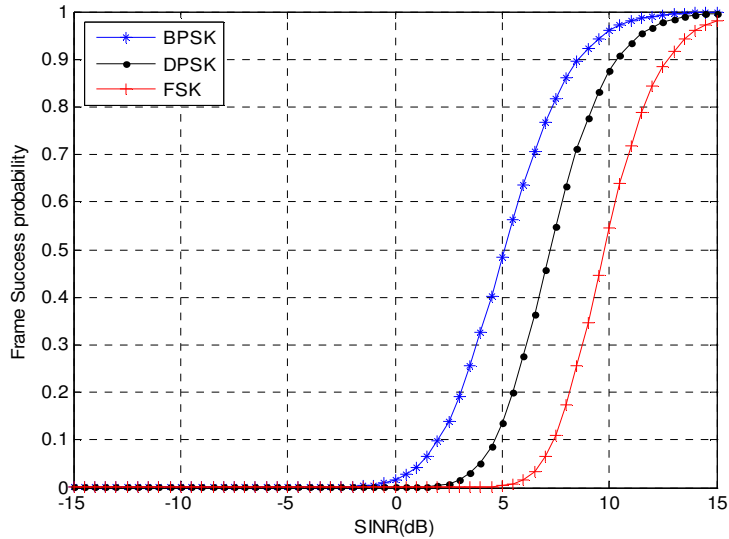

Figure 3. Probability of successful transmission

### 4.3 Power Efficiency

The performance of a modulation scheme is often measured in terms of its power efficiency. With the probability of successful transmission defined, the desired transmit power level for a link over which the packets are to be transmitted needs to be determined. Before doing so, the expected power consumption has to be considered. A scenario where a node is allowed to retransmit a packet if a transmission is unsuccessful, and it continues to retransmit until the transmission is successful is considered. Let the power level chosen by the transmitter node be P, and there are (n-1) unsuccessful transmission followed by successful transmission. The expected power efficiency for power level P is an inverse function of the expected power consumption. Then, the optimal transmit power is the power level, which will maximize the expected power efficiency.

Power efficiency describes the ability of a modulation technique to preserve the fidelity of the digital message at low power levels. Figures 4, 5 and 6 show the power efficiency attained both in the case of conventional and VMIMO scheme for different values of SINR. If SINR is low and transmitting power P is high, the power efficiency is almost zero. During worse channel conditions, a node should not transmit and if transmitting it only increases the power consumption. The result indicates that the increase in power efficiency in the case of VMIMO-WSN is due to the exploitation of multiple antennas used during transmission and reception.

International Journal Of UbiComp (IJU), Vol.1, No.3, July 2010

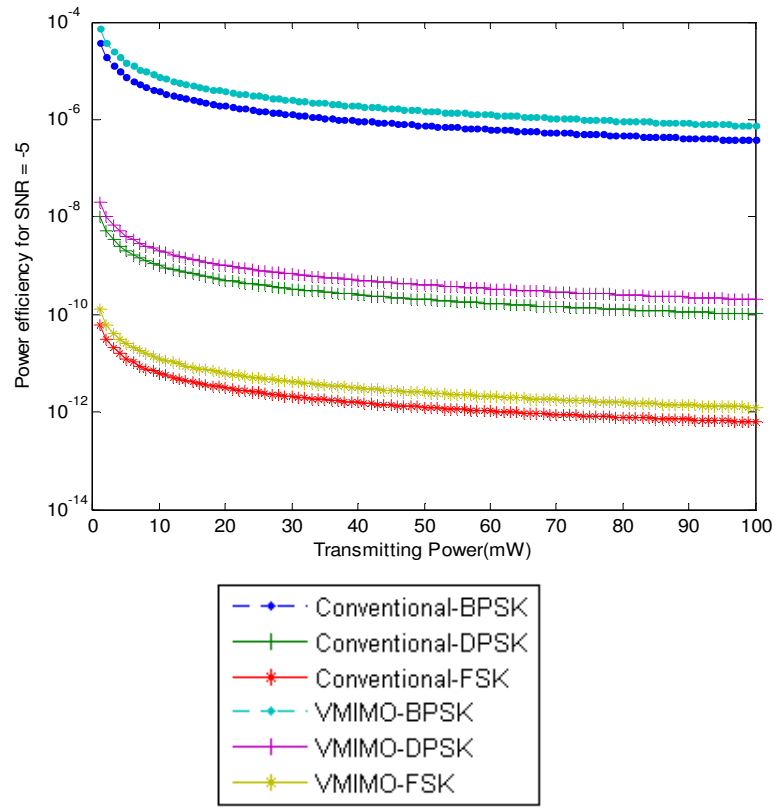

Figure 4. Power efficiency for SINR= -5dB

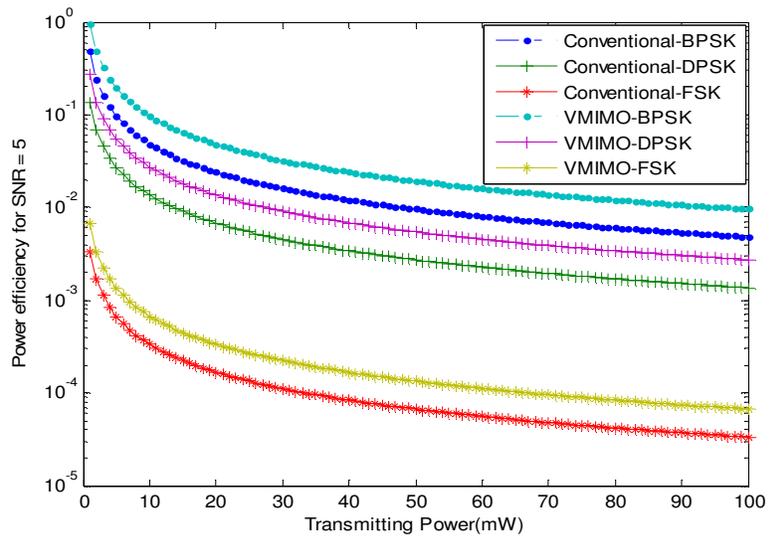

Figure 5. Power efficiency for SINR= 5dB
51

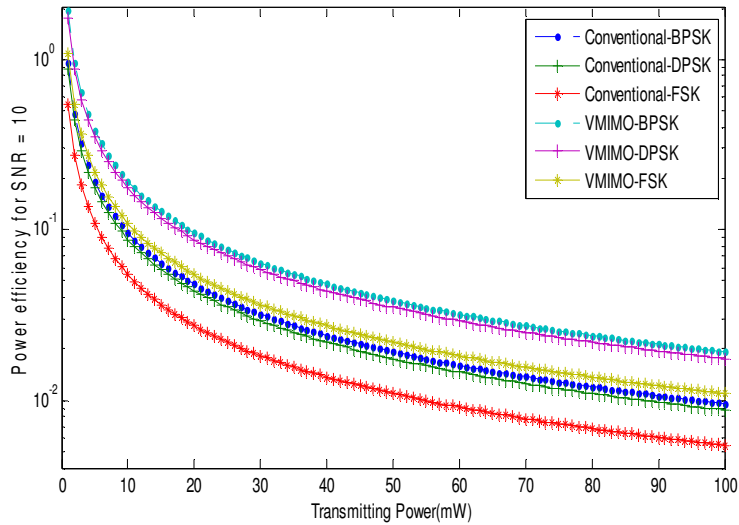

Figure 6. Power efficiency for SINR= 10 dB

At higher SINR, a node should transmit at low power to maximize its power efficiency and in this case all the modulation schemes considered provide near equal performance.

### 4.4 Net Utility

A non cooperative game model is adopted in which each node tries to maximize its net utility. Net utility is computed by considering the benefit received and the cost incurred $A(p_i)$ for transmissions as discussed in section 3.1. Figure 7 and 8 shows the disparity of the net utility with increasing transmitting power. The game is formulated such that there will be an optimal value of $p_i$, beyond which the net utility will only decline. A subset of nodes is assumed to be active and operate with fixed strategies. In VMIMO the subset of nodes cooperate to transmit the data from source to destination.

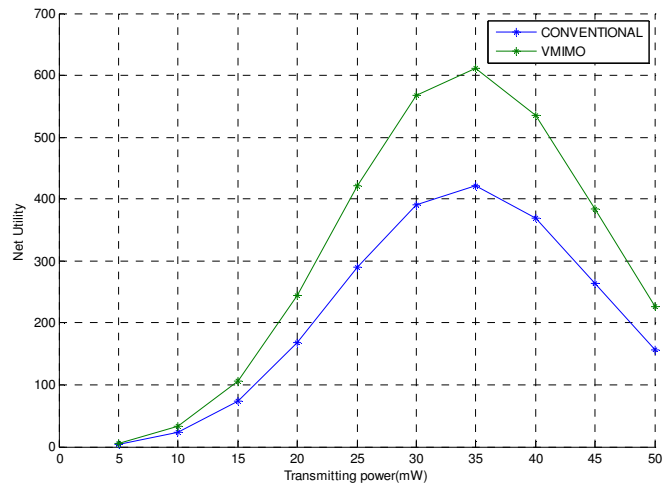

Figure 7. Net utility for uniform discrete power level

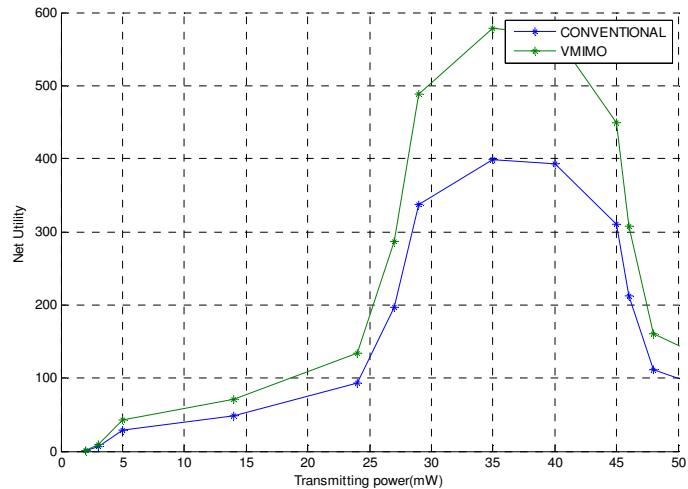

Figure 8. Net utility for non-uniform discrete power level

For any viable link, the transmitter node calculates the optimum $p_i$ ($0 < p_i \leq p_t$) such that the net utility function is maximized. When the transmitting power is 15mW, VMIMO-WSN provides an improvement of 5% in net utility as compared to the conventional scheme. As the transmitting power is increased further, at an optimal $p_i$, there is an increase in net utility by 27%. From the graph, it is intuitive that a transmitting power level of 35mW gives the best response for the node. It is also evident from the graph that even if the node unilaterally changes its strategy and does not transmit with the optimum transmitting power level, the node will not get its best response and will not be able to reach Nash equilibrium. The Nash equilibrium point is the best operating point to increase the traditional utility function.

## 5. CONCLUSIONS

This paper provides a non cooperative game theoretic approach to solve the problem of power control found in wireless sensor networks. The nodes in the sensor network cooperate to transmit the data from source to destination. A utility function with an intrinsic property of power control was designed and power allocation to nodes was built into a non-cooperative game. The performance and existence of Nash equilibrium was analyzed. In the case of VMIMO-WSN the node transmits only when the channel conditions are good and its transmission power is below the threshold power level. Results show that the game theoretic approach used in VMIMO-WSN enhances the net utility by minimizing the power at which the nodes transmit. The outcome of the simulation results also show the desired power level at which the nodes should transmit to maximize their utilities.

**Authors**

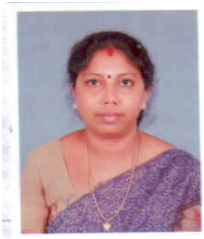
R.Valli received B.E degree in Electronics and Communication Engineering from Madras University, Chennai in 1996 and M.Tech degree in Electronics and Communication Engineering from Pondicherry Engineering College, Pondicherry in 2005. She is pursuing her Ph.D. programme in Department of Electronics and Communication Engineering, Pondicherry University. Her research interests include computer networks, wireless adhoc and sensor network.

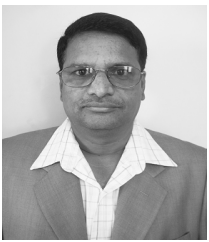
**P. Dananjayan** received Bachelor of Science from University of Madras in 1979, Bachelor of Technology in 1982 and Master of Engineering in 1984 from the Madras Institute of Technology, Chennai and Ph.D. degree from Anna University, Chennai in 1998. He is working as Professor in the Department of Electronics and Communication Engineering, Pondicherry Engineering College, Pondicherry, India. He has been as visiting professor to AIT, Bangkok. He has to his credit more than 60 publications in National and International Journals. He has presented more than 130 papers in National and International Conferences. He has guided 9 Ph.D candidates and is currently guiding 6 Ph.D students. His research interests include spread spectrum techniques, wireless communication, wireless adhoc and sensor networks.